# Electronic Single Molecule Identification of Carbohydrate Isomers by Recognition Tunneling

JongOne Im[1,3], Sovan Biswas[1,2], Hao Liu,[1,2*] Yanan Zhao[1], Suman Sen,[1,2] Sudipta Biswas[1,2], Brian Ashcroft[1], Chad Borges[1,2], Xu Wang[2], Stuart Lindsay[1,2,3] and Peiming Zhang[1]
†

**Carbohydrates, particularly those attached to proteins (glycans), play a central role as mediators in most biological processes. Examples are protein folding,[1] cell adhesion[2] and signaling,[3] fertilization[4] and embryogenesis,[5] pathogen recognition[6] and immune responses.[7] Glycan structures are complicated by isomerism. Epimers and anomers, regioisomers, and branched sequences[8] contribute to a structural variability that dwarfs those of nucleic acids and proteins. For example > 1.05 x $10^{12}$ structures are available to a hexasaccharide,[9,10] a complexity that challenges even the most sophisticated analytical tools, such as NMR and mass spectrometry. NMR is capable of determining carbohydrate structures, but it requires milligrams of sample, long data acquisition times (hours or even days), and cannot distinguish small amounts of coexisting isomers.[11] Since many carbohydrates share a molecular weight, mass spectrometry is unable to identify them without additional chemical steps.[12] The problem has recently been addressed by combining ion-mobility spectrometry, which uses collision cross-sections to separate isomers, with mass spectrometry (IM-MS),[13] but IM-MS cannot resolve closely related epimers with almost identical collision cross-sections. Emerging nanotechnologies (e.g., nanopores[14]) offer a promising alternative with the capability of detecting single-molecules. Here, we introduce an electron tunneling technique that is label-free and can identify carbohydrates at the single-molecule level, offering significant benefits over existing technology. It is capable of analyzing sub-picomole quantities of sample, counting the number of individual molecules in each subset in a population of coexisting isomers, and is quantitative over more than four orders of magnitude of concentration. It resolves epimers not well separated by ion-mobility and can be implemented on a silicon chip. It also provides a readout mechanism for direct single-molecule sequencing of linear oligosaccharides.**

---

* Current address:Beckman Coulter, 11800 SW 147th Ave, Miami, FL 33196
† [1]Biodesign Institute, [2]School of Molecular Sciences and [3]Department of Physics, Arizonan State University, Tempe, AZ 85287.

In Recognition Tunneling (RT) characteristic electron tunnel-current spikes are generated when an individual analyte is trapped by capture molecules tethered to two electrodes separated by a few nanometers. We have used RT as a single-molecule analytical tool to identify individual nucleobases, amino acids, and peptides.[15-19] In the present study of carbohydrates, the capture molecule was 4(5)-(2-mercaptoethyl)-1*H*-imidazole-2-carboxamide,[20] (ICA in Figure 1a). ICA is a molecule bearing multiple hydrogen bond donors and acceptors for recognition and a two-carbon alkyl chain terminated with a thiol function for attachment to electrodes. We first applied RT to distinguish between two anomeric isomers, namely methyl α-D-glucopyranoside (α-$^{M}$Glu) and methyl β-D-glucopyranoside (β-$^{M}$Glu). They are non-reducing monosaccharides, differing only in the relative orientation of the methoxy group (marked in red), as shown in Figs 1b and c. We acquired these materials from a commercial source (Section 1 of Supplementary Information, hereafter referred to as SI) and confirmed their configurations by NOESY NMR before use (Fig. S1, SI). Theoretical simulations show that each of them can form a hydrogen-bonded triplet with a pair of ICA molecules respectively in a 2.2 nm wide tunnel gap, having a distinguishable hydrogen-bonding pattern and energy (Figs. 1d and e). α-$^{M}$Glu forms two hydrogen bonds with each ICA, resulting in a more stable complex than that of β-$^{M}$Glu, which interacts with each ICA through a single hydrogen bond. The formation of 1:1 and 2:1 complexes between ICA and these two anomers in aqueous solution was detected by ESI-MS (Tables S3 and S4, SI; MS data for the other monosaccharides and disaccharides used in the present work are also included). RT measurements followed a process of functionalizing Pd-STM probes and Pd-substrates with ICA in ethanolic solutions (see details in Section 6, SI), mounting them in a PicoSPM scanning tunneling microscope (Agilent), introducing an analyte solution (typically 100 μM in 1 mM phosphate buffer, pH 7.4) to the liquid cell, and collecting current recordings with a tip-substrate bias of 0.5 V (see experimental details in Section 7, SI). Examples of the tunnel current signals generated by $\alpha$–$^{M}$Glu and the $\beta$–$^{M}$Glu are shown in Figs 1f and g (control experiments in Section 7.1, SI). They consist of a series of current spikes similar to those theoretically predicted for other hydrogen-bonded structures.[21] Driven by thermal fluctuations, the signals are stochastic but rich in information about the bonding, and thus the identity, of the trapped analyte molecule. Individual signal features have broad and overlapping distributions, as shown for the two anomers by the green and red curves in Figs 1h and i. However, when a three-dimensional probability density plot is made of the *pairs* of values of signal features that occur together in each signal spike (or cluster of spikes), the distributions can be separated quite well. This is illustrated in Figure 1j, which uses the two signal features that were plotted separately in Figs 1h and I, but with the additional information about how often particular pairs of values occurred together. On this plot the brightness indicates the third dimension, i.e., frequency, with the red points showing data for $\alpha$–$^{M}$Glu and the green points showing data for $\beta$–$^{M}$Glu. Colors are mixed so that overlapped regions appear as yellow. The 3D distributions are quite complicated with many “islands” of clustered data but the distributions for the two analytes are well separated enough that a new and unassigned data point can be classified as one of the two anomers with 80$\pm$5% accuracy. The two signal features used in the figures were chosen in order to demonstrate the underlying mechanism of the much more accurate analysis that can be achieved by including many more signal features using a Support Vector Machine, SVM, a machine-learning algorithm,[19] the details of which are described elsewhere.[22]

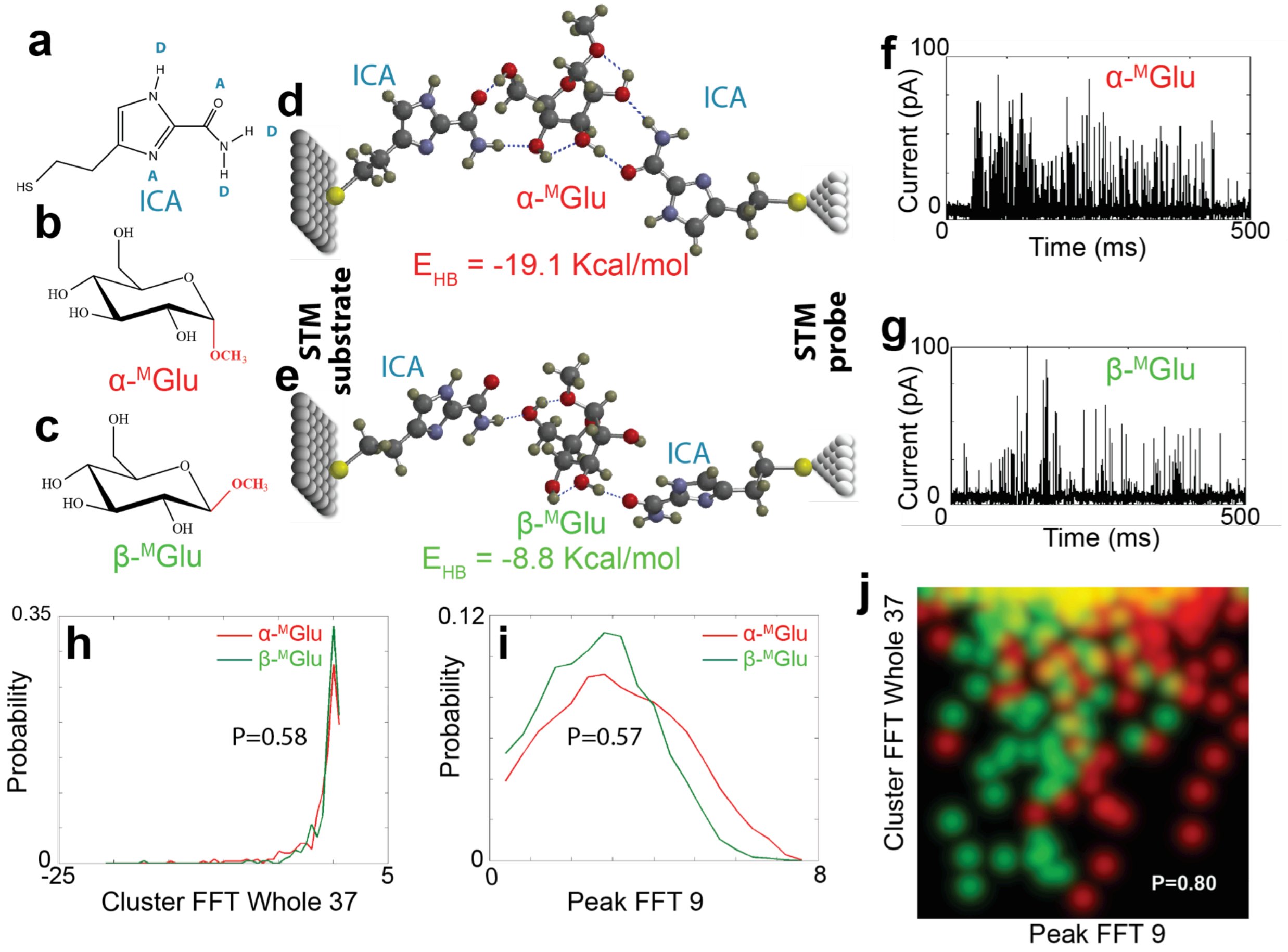

**Figure 1:** RT analysis of anomers of methyl D-glucopyranoside. (a) The recognition molecule ICA contains a thiol linkage to bond to metal electrodes, as well as a number of hydrogen bonding donors (D) and acceptors (A) through which a large range of analytes can be can captured by a diversity of spatial arrangements resulting from tautomerism and rotation about σ-bonds; (b) Structure of α-$^{M}$Glu and (c) Structure of β-$^{M}$Glu, both of which can form hydrogen-bonded triplets with ICA molecules spanning a tunnel gap of 2.2 nm, as shown by simulations in (d) and (e). Evidence of these complexes is provided by the current-spikes that appear only after an analyte solution is added to pure buffer solution in a tunnel gap (f and g). Distributions of signal features are broad and overlapped (red = α-$^{M}$Glu, green = β-$^{M}$Glu) as shown here for one frequency band in the Fourier transform of signal clusters (cluster FFT whole 37 –h) and for a band in the Fourier transform of individual peaks (Peak FFT 9 –i). Data can only be assigned to one analye or the other with a probability only marginally above random, P=0.5 (see Methods for details of the signal analysis). However, when the same two distributions are plotted together in a 3D histogram (j) where the brightness of each point represents the frequency with which a particular *pair* of values occur, the accuracy with which data can be assigned increases to 80%. This accuracy can be improved to ~ 99% using additional signal features. Colors in (j) are mixed so overlapped points are yellow.

In brief, an SVM was first trained using a randomly-selected 10% subset of

**Table 1 | Accuracy of determining carbohydrate pairs by SVM analysis of RT data**

| Analyte | Accuracy (%) | Analyte | Accuracy (%) |
|---|---|---|---|
| 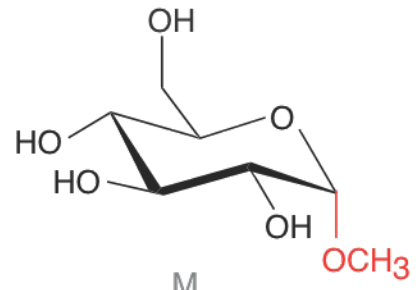 D-α-$^{M}$Glu | 99.1 | 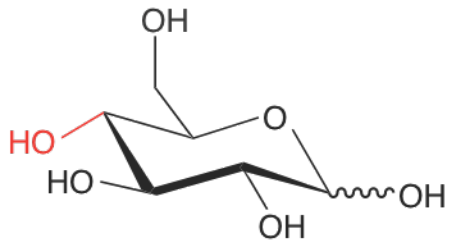 D-Glucose | 99.7 |
| 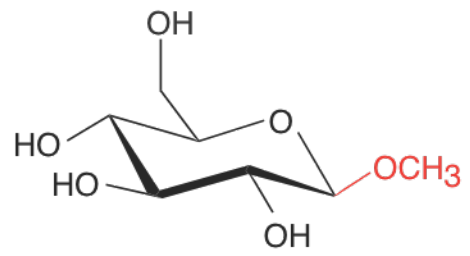 D-β-$^{M}$Glu | 99.4 | 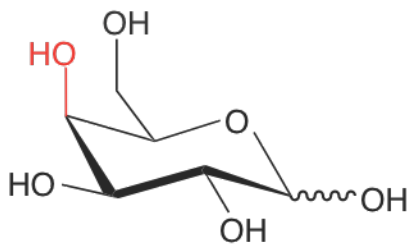 D-Galactose | 99.5 |
| 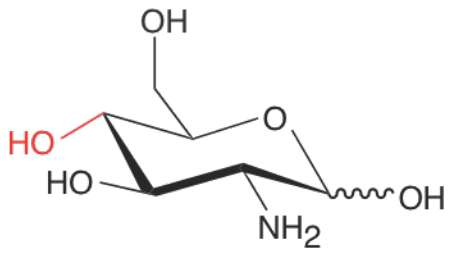 D-Glucosamine | 98.9 | 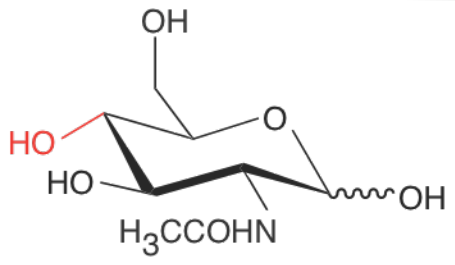 D-N-Acetyl-Glucosamine | 99.3 |
| 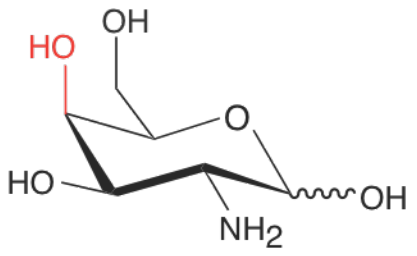 D-Galactosamine | 98.7 | 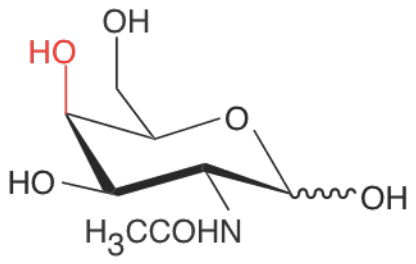 D-N-Acetyl-Galactosamine | 99.5 |
| 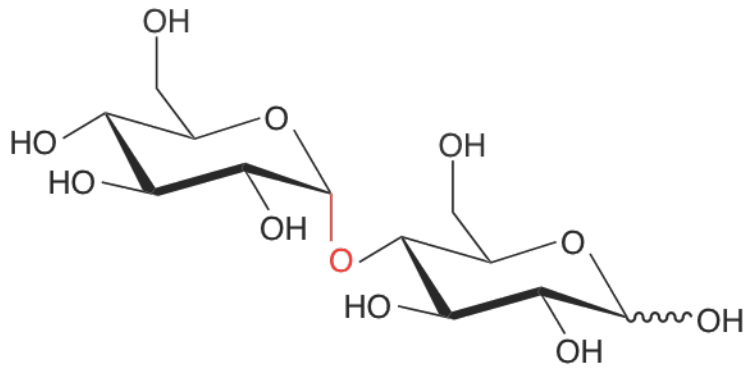 α-D-Glucopyranosyl-(1→4)-D-glucopyranose (Maltose) | 99.5 | 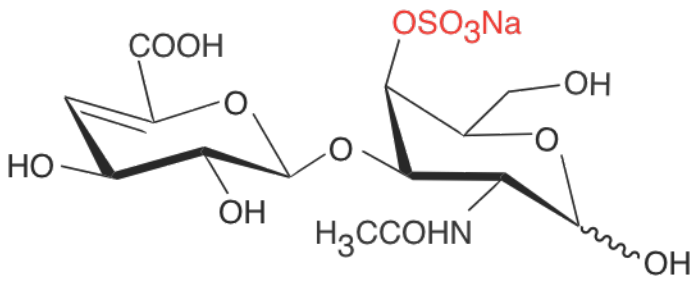 β-Δ$^{4,5}$-D-UA(1→3)-D-GalNAc-4-O-Sulfate (D0A4)* | 98.8 |
| 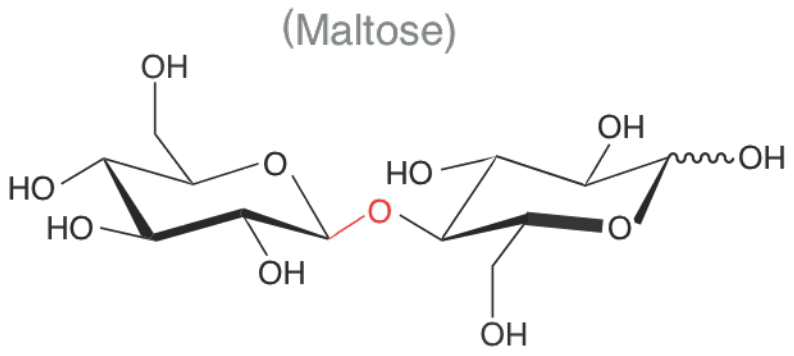 β-D-glucopyranosyl-(1→4)-D-glucopyranose (Cellobiose) | 99.8 | 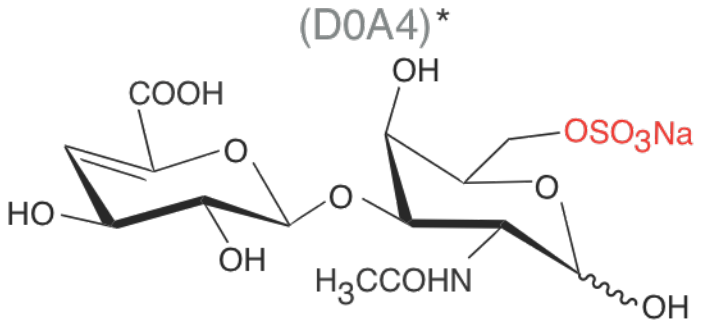 β-Δ$^{4,5}$-D-UA(1→3)-D-GalNAc-6-O-Sulfate (D0A6)* | 99.6 |

* Lawrence disaccharide nomenclature used to indicate sulfation position[23]

data from known analytes, based on up to 264 available signal features such as amplitude, pulse shape, Fourier components, and so on. [22] This feature set is iteratively reduced to leave the smallest number of signal features that give a high accuracy for classifying the data, as tested on the remaining 90% of the data not used in training. The SVM first identified events that were common to data obtained from different samples owing to contamination, capture events that were insensitive to chemical variation and noise spikes generated by the STM electronics and servo control. These amounted to about 50% of all signal spikes. The remaining signals could be classified with very high accuracies (> 99% for the two anomers - first entry, Table 1). With similar RT measurements, we were able to distinguish glucose, glucosamine, and N-acetylgluosamine from their respective C-4 epimers with ~ 99% accuracy (Table 1). Note that these monosaccharides are reducing sugars so that each of them exists with an equilibrium distribution of anomeric isomers and a possible open-chain form in aqueous solution. Nonetheless, the identification of these epimers was accurate, because the training set for each epimer contained the same mixture of anomers. We were also able to distinguish accurately between two anomeric isomers of a disaccharide, maltose ($\alpha$-D-glucopyranosyl-(1→4)-D-glucopyranose) and cellobiose ($\beta$-D-glucopyranosyl-(1→4)-D-glucopyranose) (Table 1), and between two regioisomers, D0A4 and D0A6, of a chondroitin sulfate (synthesis is described in Section 2, SI), a repeating disaccharide unit of glycosaminoglycans (GAGs) (Table 1).

**Table 2 | Accuracy of determining individual carbohydrates from a pool of RT data**

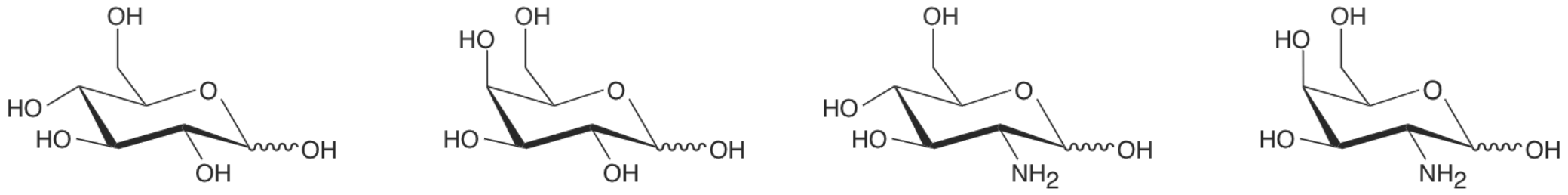

| D-Glucose | D-Galactose | D-Glucosamine | D-Galactosamine |
|---|---|---|---|
| 97.7 | 96.3 | 94.7 | 98.1 |

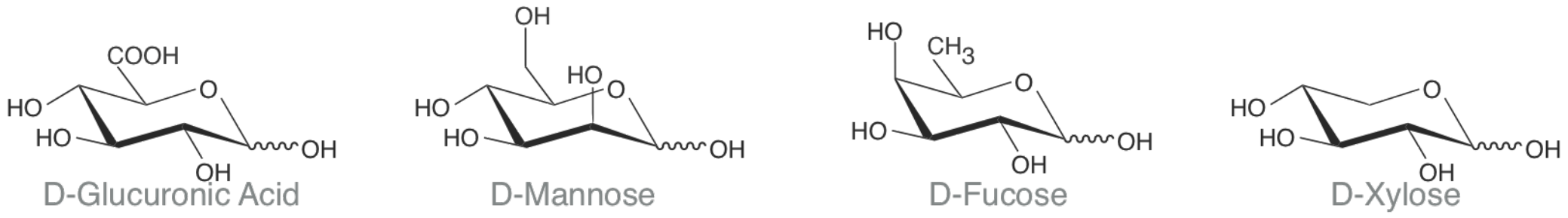

| D-Glucuronic Acid | D-Mannose | D-Fucose | D-Xylose |
|---|---|---|---|
| 96.1 | 94.3 | 96.7 | 98.1 |

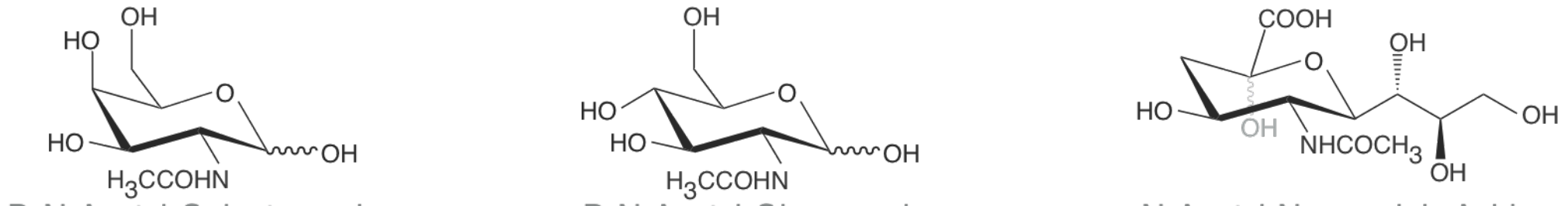

| D-N-Acetyl-Galactosamine | D-N-Acetyl-Glucosamine | N-Acetyl-Neuraminic Acid |
|---|---|---|
| 97.5 | 96.0 | 97.6 |

Red number indicates the accuracy in percentage.

Strikingly, RT was capable of identifying many different monosaccharides from pooled data with a high degree of accuracy. This is illustrated in Table 2, which lists the average accuracy with which each single-signal spike was assigned. These samples included D-glucosamine, D-galactosamine, and D-mannose, abundant in the mammalian glycome [24,25] as well as N-acetyl-neuraminic acid, the predominant sialic acid found in mammalian cells. Accuracies were in excess of 94% (random would be 9%). In contrast, IM-MS is ineffective in discriminating between galactose and mannose [26] or between glucose and galactose.[13]

We have determined the binding affinity of ICA to α-$^{M}$Glu in the tunnel gap via a serial dilution study. Figure 2 is a plot of normalized signal peak frequency vs. sample concentration where the error bars represent the standard deviation of 3 to 5 repeated measurements (Table S7). The data were well fitted by a Langmuir function ($R^2$ = 0.976) from which a dissociation constant, $K_d$ = 0.74 ± 0.25 μM was obtained (insert in Figure 2, and also see Section 8 of SI for more details). This represents a striking improvement on the affinity of a single ICA molecule for this sugar molecule. We used surface plasmon resonance (SPR) to measure $K_d$ for the adsorption of α-$^{M}$Glu on an ICA monolayer, finding a value of 4 mM (Section 9 of SI, Table S8). One important contribution to the enhanced affinity comes from the simultaneous interaction of the sugar molecule with a fixed pair of ICA in the tunnel gap. Assuming that entropy changes were same for both ICA molecules, and equal to that for a single binding event, the adsorption free energy would be doubled on binding at two sites. Consequently, $K_d$ for the two-site binding can be as large as the square of the value for binding by one site, i.e. (4 mM)$^2$ or 16 μM. This is still 20 times larger than the observed 0.74 μM. The electric field in the gap exceeds $10^8$ V/m and may contribute to this enhancement by increasing retention of bound molecules. The additional free energy owing to the molecular dipole (calculated to be 1.17 Debye for α-$^{M}$Glu) can be up to 20% of thermal energy and this will extend the retention of trapped molecules. The same field may also result in dielectrophoretic concentration of molecules in the vicinity of the junction, kinetically enhancing the effective on-rate. These factors combined lead to a remarkably enhanced capture affinity. The count rate at a given concentration is quite reproducible (Table S7) and Figure 2 shows that the RT measurement has a dynamic range that is greater than 4 orders of magnitude. Significant count rates are obtained with concentrations as low as 10 to 100 nM. This quantitative ability offers a way to count the number of signal spikes of a given character so as to quantify the relative amount of a given isomer. In comparison, mass spectroscopy, which is not inherently quantitative, requires additional techniques, such as isotope labeling for quantification of sample concentrations. The present experiments used a large volume (200 μL) of sample to fill the liquid cell of the scanning tunneling microscope. However, micron-scale solid-state tunnel junctions are becoming available[27] and it is practical to use microliter volumes with such small devices. With microliter volumes and 10 to 100 nM concentrations, only $10^{-13}$ to $10^{-14}$ moles of sample will be required for RT measurements in the future.

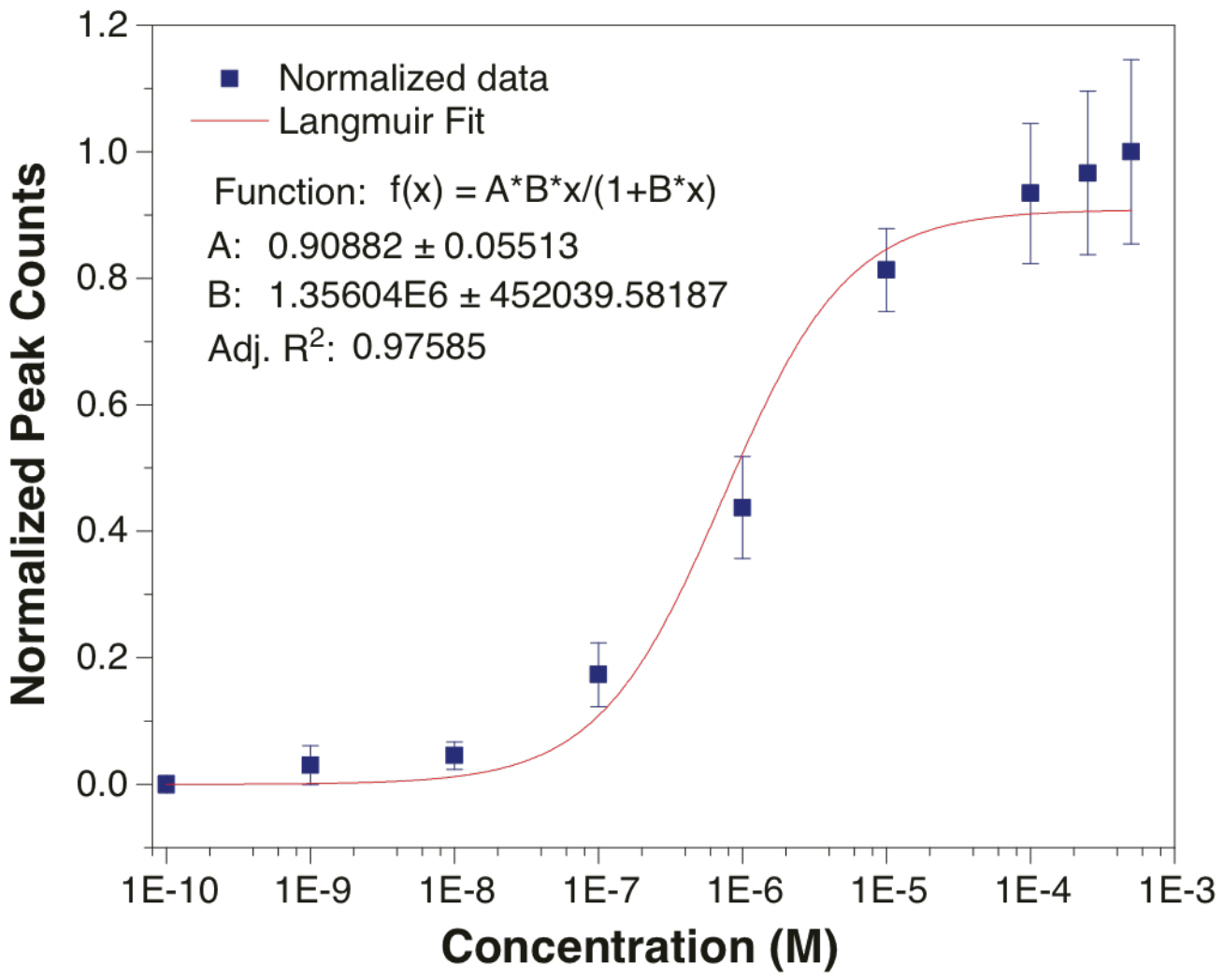

**Figure 2:** Plot of normalized RT counting rates vs concentration of α-$^{M}$Glu for trapping the analyte in an RT gap functionalized with ICA molecules and a fit to a Langmuir isotherm.

In the present work, we demonstrated the power of RT to discriminate between individual saccharides. As is the case for IM-MS,[13] identifying a broad range of analytes will require construction of the appropriate databases, and this will prove to be a more demanding task for RT because IM-MS also collects independent measurements of ion-masses, providing an independent check of signal assignments. Clearly, neither technique could analyze the vast number of possible isomers in a long oligosaccharide.[9] However, RT is clearly capable of providing a key component of a single-molecule sequencing system for linear oligosaccharides, a presently almost intractable problem. It has been proposed that RT electrodes incorporated into a nanopore can be used for sequencing linear heteropolymers such as DNA.[28] The present work shows that RT can identify a large number of individual glycans accurately, so that, combined with a nanopore device to present the individual sugars to the electrodes sequentially, the compositional sequence of linear oligosaccharides might be read directly. Glycosaminoglycans represent an example of an important class of linear oligosaccharide for which the compositional sequence is very difficult to obtain by current methods, but with a compositional variability that is small enough to make use of a data-base practical for sequencing.[29] Thus, RT may prove to be an important adjunct to IM-MS with a unique capability for analysis of very small amounts of sample and also enabling the sequence of linear oligosaccharides to be read directly.

1 Parodi, A. J. Protein glucosylation and its role in protein folding. *Annu. Rev. Biochem.* **69**, 69-93 (2000).
2 Zhao, Y. Y. *et al.* Functional roles of N-glycans in cell signaling and cell adhesion in cancer. *Cancer sci.* **99**, 1304-1310 (2008).
3 Ohtsubo, K. & Marth, J. D. Glycosylation in cellular mechanisms of health and disease. *Cell* **126**, 855-867 (2006).
4 Rosati, F., Capone, A., Giovampaola, C. D., Brettoni, C. & Focarelli, R. Sperm-egg interaction at fertilization: glycans as recognition signals. *Int. J. Dev.Biol.* **44**, 609-618 (2000).
5 Lowe, J. B. & Marth, J. D. A genetic approach to Mammalian glycan function. *Annu. Rev. Biochem.* **72**, 643-691 (2003).

6 Kawai, T. & Akira, S. The roles of TLRs, RLRs and NLRs in pathogen recognition. *Int. Immunol*. **21**, 317-337 (2009).
7 Zhang, X. L. Roles of glycans and glycopeptides in immune system and immune-related diseases. *Curr. Med. Chem.* **13**, 1141-1147 (2006).
8 Bertozzi, C. R. & Rabuka, D. in *Essentials of Glycobilogy* (Second Edition, eds Ajit Varki *et al.*) Ch. 2, 23-36 (Cold Spring Harbor Laboratory Pess, New York, 2009).
9 Laine, R. A. A calculation of all possible oligosaccharide isomers both branched and linear yields 1.05 x 10(12) structures for a reducing hexasaccharide: the Isomer Barrier to development of single-method saccharide sequencing or synthesis systems. *Glycobiology* **4**, 759-767 (1994).
10 Han, X., Zheng, Y., Munro, C. J., Ji, Y. & Braunschweig, A. B. Carbohydrate nanotechnology: hierarchical assembly using nature's other information carrying biopolymers. *Curr. Opin. Biotechnol.* **34**, 41-47 (2015).
11 Duus, J., Gotfredsen, C. H. & Bock, K. Carbohydrate structural determination by NMR spectroscopy: modern methods and limitations. *Chem. Rev.* **100**, 4589-4614 (2000).
12 Nagy, G. & Pohl, N. L. Monosaccharide identification as a first step toward de novo carbohydrate sequencing: mass spectrometry strategy for the identification and differentiation of diastereomeric and enantiomeric pentose isomers. *Anal. Chem.* **87**, 4566-4571 (2015).
13 Hofmann, J., Hahm, H. S., Seeberger, P. H. & Pagel, K. Identification of carbohydrate anomers using ion mobility-mass spectrometry. *Nature* **526**, 241-244 (2015).
14 Fennouri, A. *et al.* Single molecule detection of glycosaminoglycan hyaluronic acid oligosaccharides and depolymerization enzyme activity using a protein nanopore. *ACS nano* **6**, 9672–9678 (2012).
15 Chang, S. *et al.* Electronic signature of all four DNA nucleosides in a tunneling gap. *Nano Lett.* **10**, 1070-1075 (2010).
16 Huang, S. *et al.* Recognition tunneling measurement of the conductance of DNA bases embedded in self-assembled monolayers. *J. Phys. Chem. C* **114**, 20443–22044 (2010).
17 Huang, S. *et al.* Identifying single bases in a DNA oligomer with electron tunneling. *Nat. Nanotechnol.* **5**, 868-873 (2010).
18 Lindsay, S. *et al.* Recognition tunneling. *Nanotechnology* **21**, 262001-262013 (2010).
19 Zhao, Y. *et al.* Single molecule spectroscopy of amino acids and peptides by recognition tunneling. *Nat. Nanotechnol.* **9**, 466-473 (2014).
20 Liang, F., Li, S., Lindsay, S. & Zhang, P. Synthesis, physicochemicalpProperties, and hydrogen bonding of 4(5)-substituted-1*H*-imidazole-2-carboxamide, a potential universal reader for DNA sequencing by recognition tunneling. *Chem. - Eur. J.***18**, 5998 – 6007 (2012).
21 Krstić, P., Ashcroft, B. & Lindsay, S. Physical model for recognition tunneling. *Nanotechnology* **26**, 084001 (2015).
22 Chang, S. *et al.* Chemical recognition and binding kinetics in a functionalized tunnel junction. *Nanotechnology* **23**, 235101 (2012).
23 Lawrence, R., Lu, H., Rosenberg, R. D., Esko, J. D. & Zhang, L. Disaccharide structure code for the easy representation of constituent oligosaccharides from glycosaminoglycans. *Nat. Methods* **5**, 291-292 (2008).
24 Werz, D. B. *et al.* Exploring the Structural Diversity of Mammalian Carbohydrates (“Glycospace”) by Statistical Databank Analysis. *ACS Chem. Biol.* **2**, 685–691 (2007).
25 Adibekian, A. *et al.* Comparative bioinformatics analysis of the mammalian and bacterial glycomes. *Chem. Sci.* **2**, 337-344 (2011).
26 Both, P. *et al.* Discrimination of epimeric glycans and glycopeptides using IM-MS and its potential for carbohydrate sequencing. *Nat. Chem.* **6**, 65-74 (2014).
27 Pang, P. *et al.* Fixed Gap Tunnel Junction for Reading DNA Nucleotides. *ACS Nano* **8**, 11994–12003 (2014).
28 Branton, D. *et al.* Nanopore Sequencing. *Nat. Biotechnol.* **26**, 1146-1153 (2008).
29 Prydz, K. Determinants of Glycosaminoglycan (GAG) Structure. *Biomolecules* **5**, 2003-2022 (2015).

**Supplementary Information** is available in the online version of the paper.

**Acknowledgment** This work was supported by grant HG006323 from the National Human Genome Research Institute and R21GM118339 from National Institute of General Medical Sciences.

**Author Contributions** J.I., H.L., and Y.Z. carried out RT measurements; J.I. and B.A. and S.L. RT data analysis; So.B. synthesis of ICA, NMR and MS experiments and data analysis; Su.S. characterization of functionalized Pd surfaces; Su.B. and P.Z. SPR experiments and data analysis; X.W. synthesis of D0A4 and D0A6; P.Z. computer modeling; C.B. designed MS experiments and data analysis; S.L. and P.Z. designed the project and wrote the manuscript.

**Author Information** Reprints and permissions information is available at www.nature.com/reprints. S.L. and P.Z. are named as inventors in patent applications and SL is cofounder of Recognition AnalytiX, a company that has licensed RT technology in the field of glycomics. Readers are welcome to comment on the online version of the paper. Correspondence and requests for materials should be addressed to S.L. (stuart.lindsay@asu.edu) or P.Z (peiming.zhang@asu.edu).

# SUPPLEMENTARY INFORMATION

## Table of Contents

## 1. Chemicals and reagents

Monosaccharides, maltose, and cellobiose were purchased from Sigma-Aldrich (purity ≥ 99%). 4-O-sulfated-chondroitin sulfate disaccharides and 6-O-sulfated chondroitin sulfate disaccharides (D0A4 and D0A6) were synthesized in-house. Water was purified in a Milli-Q system with a resistivity of ~18.2 MΩ.cm and total organic carbon of less than 5 ppb. Sample solutions (100 μM) were prepared in a sodium phosphate buffer (1 mM, pH 7.4). All the solutions were freshly prepared right before measurements. NMR spectra were recorded at 400 MHz ($^1$H) and 100 MHz ($^{13}$C), respectively. Chemical shifts are given in parts per million (ppm) on the delta scale (δ) and referred either to tetramethylsilane ($^1$H and $^{13}$C NMR δ = 0 ppm) or the residual solvent peak.

## 2. Preparation of D0A4 and D0A6

Disaccharides D0A4 and D0A6 were obtained by enzymatic digestion of native CS-A polysaccharides using chondroitinase ABC. Specifically, CS-A (0.1 g, Sigma Aldrich) was digested using chondroitinase ABC enzyme (1 U, Sigma Aldrich) in a digestion buffer (50 mM Tris pH 8.0, 60 mM sodium Acetate, 12 mL) at 37 °C for 24 hours. Disaccharides were separated from other larger fragments using a 2.5 cm × 175 cm size exclusion chromatography column (Bio-Rad Biogel P10 resin). Fractions containing disaccharides were combined and desalted. D0A4 was then separated from D0A6 using analytical strong anion exchange (SAX) HPLC (Waters Spherisorb SAX analytical column) with a salt gradient of 0 to 1 M NaCl at pH 4.0. The structures of the disaccharides were confirmed using $^{13}$C-edited heteronuclear single quantum coherence (HSQC) NMR spectroscopy (Table S1). ESI-MS analysis for D0A4: found m/z 482.05 (M + H), calcd for $C_{14}H_{20}NNaO_{14}S$ 481.0502; for D0A6: found m/z 482.06 (M + H), calcd for $C_{14}H_{20}NNaO_{14}S$ 481.0502.

**Table S1.** Chemical shift assignments for D0A4 and D0A6

| | D0A4 | | D0A6 | |
|---|---|---|---|---|
| | $^{13}$C | $^1$H | $^{13}$C | $^1$H |
| GluA.1 | 102.6 | 5.29, 5.26 | 104 | 5.26, 5.21 |
| GluA.2 | 71.2 | 3.83 | 72.2 | 3.82 |
| GluA.3 | 67.2 | 3.94 | 69 | 4.13 |
| GluA.4 | 109.3 | 5.96 | 110.1 | 5.91 |
| GalNAc.1 | 93.8, 97.5 | 5.20, 4.76 | 94.1, 97.8 | 5.24, 4.75 |
| GalNAc.2 | 52.4, 55.9 | 4.36, 4.05 | 51.6, 55 | 4.31, 4.02 |
| GalNAc.3 | 75.6, 78.3 | 4.29, 4.15 | 79.5, 82.3 | 4.15, 3.96 |
| GalNAc.4 | 80.1, 78.7 | 4.67, 4.60 | 71.0, 70.2 | 4.25, 4.18 |
| GalNAc.5 | 73.1, 77.3 | 4.27, 3.84 | 71.2, 75.5 | 4.39, 3.98 |
| GalNAc.6 | 63.8 | 3.77, 3.69 | 70.9 | 4.22, 4.15 |

## 3. NOESY NMR of α-$^M$Glu and β-$^M$Glu

300 mM stock solutions of β-$^M$Glu and α-$^M$Glu were prepared in DMSO-$d_6$ at room temperature under inert atmosphere. A volume of 0.75 mL solution from each stock solution was used for an individual NMR experiment. $^1$H, COSY and NOESY were recorded in a Varian 500 MHz NMR at 25°C at the Arizona State University NMR facility lab. The NOE mixing time was set to 400 ms and a total number of scans were set to 16. Each spectrum was recorded for 3 h at 25°C. Data was plotted in VnmrJ 4.0 and exported to adobe illustrator. In case of the α-anomer, a cross peak between $H_1$ and $H_2$ was observed (Fig. S1, a) as expected. For the β-anomer, two NOE cross peaks were observed between $H_1$ and $H_3$ as well as between $H_1$ and $H_5$ protons (Fig. S1, b).

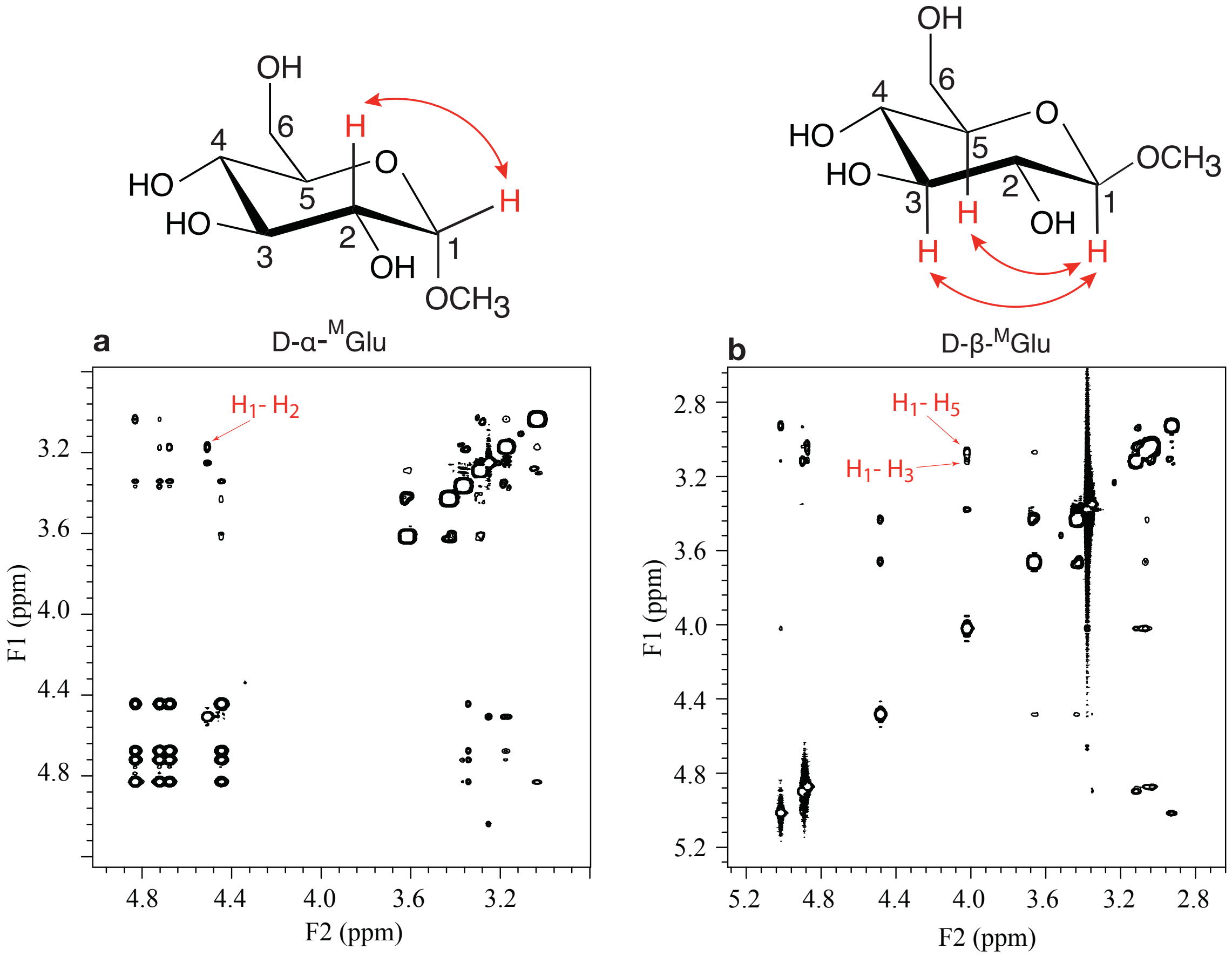

**Figure 1.** $^1H$–$^1H$ NOESY NMR spectra of (a) α-$^M$Glu and (b) β-$^M$Glu

## 4. Computation of ICA triplets with α-$^M$Glu and β-$^M$Glu.

DFT calculations were performed using the program Spartan'14 for Windows, Wavefunction Inc. Individual 2D chemical structures of β-$^M$Glu, α-$^M$Glu and ICA were drawn in ChemBioDraw Ultra 14.0, and exported to Spartan'14 to generate the corresponding 3D structures, where hydrogen bonded triplexes were formed by energy minimization using the built-in MMFF94s molecular mechanics. The complexed structures were further optimized using B3LYP/6-31G* in water with two sulfurs constrained at a distance of 2.2 nanometer. The final results are shown in Figure 1d and e.

## 5. Complexes of ICA with carbohydrates measured by Electrospray ionization mass spectrometry (ESI-MS).

ESI can preserve non-covalent complexes in going from their native solution state into the gas phase in the form of single or multiple charged ions.[1] In the present study we observed stoichiometric complexes between ICA and carbohydrates. By means of ESI-MS, we first obtained characteristic m/z peaks for each individual carbohydrate and ICA in aqueous solution, and then measured the aqueous solutions of ICA mixed with each carbohydrate in 2:1 ratio. Data showed that all monosaccharides and ICA existed as self-associated dimers in solution along with their monomeric form (Table S2), ICA formed 1:1 molecular complex ions with all carbohydrates (Table S3), and 2:1 molecular complex ions with most of the carbohydrates (Table S4). Each molecular ion complex was confirmed by tandem mass spectrometry (MS/MS) and reported as MS/MS product ion in the respective columns.

**Experimental details.** ICA (200 μM) and carbohydrate (100 μM each) solutions were respectively prepared in water and sparged with argon. Each sample solution was injected into a Bruker maXis 4G electrospray ionization quadrupole time-of-flight (ESI-Q-TOF) mass spectrometer at a 3 μL/min infusion rate via syringe pump. Tandem (MS/MS) mass spectrometry was used to observe product ion peaks from molecular complex ion peaks to confirm the composition of the molecular complex. The ESI source was equipped with a microflow nebulizer needle operated in a positive ion mode. The spray needle was held at ground and the inlet capillary set to -4500 V. The end plate offset was set to -500 V. The nebulizer gas and dry gas ($N_2$) were set to 1.2 Bar and 1.5 L/min, respectively, and the dry gas was heated to 220°C. In TOF-only mode the quadrupole ion energy was set to 4 eV and the collision energy was set to 1 eV. Collision gas (Ar) was set to a flow rate of 20%. In most cases MS/MS experiments were conducted with a precursor ion isolation width of 3 *m/z* units. However, if other ions were present in this range precursor ion isolation width was set to 1 *m/z* unit. Collision energy was set to 10-20 eV, which was sufficient to fragment non-covalent complexes. Each spectrum was recorded over a time period of 0.5 to 1 min. Typically a spectrum acquired for one minute is an accumulation of 60 separate recorded mass spectra averaged across 1 min time period. Signal to noise ratio greater than three (S/N>3) was used to define the limit of detection. Due to the lack of an acid modifier in the infused solutions, most carbohydrates and molecular complexes were observed as single or multiply sodium ions $[M+nNa-(n-1)H]^+$ rather than as protonated molecular form $[M + H]^+$. Average mass accuracy was within 0.025 Da.

**Table S2.** MS data of individual Carbohydrates and ICA

| Analytes | Calculated Monoisotopic Mass | [1]Observed *m/z* | Analytes | Calculated Monoisotopic Mass | [1]Observed *m/z* |
|---|---|---|---|---|---|
| Galactose ($C_6H_{12}O_6$) | 180.0634 | 203.05, $[M+Na]^+$, (74)<br>225.03, $[M+2Na-H]^+$, (2)<br>383.12, $[2M+Na]^+$, (100)<br>405.10, $[2M+2Na-H]^+$, (2) | α-$^M$Glu ($C_7H_{14}O_6$) | 194.0790 | 217.07, $[M+Na]^+$, (67)<br>239.05, $[M+2Na-H]^+$, (2)<br>411.15, $[2M+Na]^+$, (100)<br>433.13, $[2M+2Na-H]^+$, (1) |
| Glucose ($C_6H_{12}O_6$) | 180.0634 | 203.05, $[M+Na]^+$, (87)<br>225.03, $[M+2Na-H]^+$, (2)<br>383.12, $[2M+Na]^+$, (100)<br>405.10, $[2M+2Na-H]^+$, (3) | β-$^M$Glu ($C_7H_{14}O_6$) | 194.0790 | 217.07, $[M+Na]^+$, (68)<br>239.05, $[M+2Na-H]^+$, (2)<br>411.15, $[2M+Na]^+$, (100)<br>433.13, $[2M+2Na-H]^+$, (1) |
| Galactosamine ($C_6H_{13}NO_5$) | 179.0794 | 180.09, $[M+H]^+$, (100)<br>202.07, $[M+Na]^+$, (16)<br>359.17, $[2M+H]^+$, (64)<br>381.15, $[2M+Na]^+$, (19)<br>162.08, $[M-H_2O+H]^+$, (46) | Maltose ($C_{12}H_{22}O_{11}$) | 342.1162 | 365.11, $[M+Na]^+$, (100)<br>685.24, $[2M+H]^+$, (0.2)<br>707.22, $[2M+Na]^+$, (36)<br>729.21, $[2M+2Na-H]^+$, (0.1) |
| Glucosamine ($C_6H_{13}NO_5$) | 179.0794 | 180.09, $[M+H]^+$, (100)<br>202.07, $[M+Na]^+$, (18)<br>359.17, $[2M+H]^+$, (49)<br>381.15, $[2M+Na]^+$, (23)<br>162.08, $[M-H_2O+H]^+$, (33) | Cellobiose ($C_{12}H_{22}O_{11}$) | 342.1162 | 343.13, $[M+H]^+$, (0.3)<br>365.11, $[M+Na]^+$, (100)<br>685.24, $[2M+H]^+$, (2.2)<br>707.22, $[2M+Na]^+$, (38) |
| *N*-Acetyl-Galactosamine ($C_8H_{15}NO_6$) | 221.0899 | 222.10, $[M+H]^+$, (0.01)<br>244.08, $[M+Na]^+$, (71)<br>266.06, $[M+2Na-H]^+$, (1)<br>465.17, $[2M+Na]^+$, (100)<br>487.15, $[2M+2Na-H]^+$, (1) | Xylose ($C_5H_{10}O_5$) | 150.0528 | 173.04, $[M+Na]^+$, (100)<br>195.02, $[M+2Na-H]^+$, (2)<br>323.09, $[2M+Na]^+$, (81)<br>345.07, $[2M+2Na-H]^+$, (3)<br>367.06, $[2M+3Na-2H]^+$, (0.1) |
| *N*-Acetyl-Glucosamine ($C_8H_{15}NO_6$) | 221.0899 | 222.10, $[M+H]^+$, (1)<br>244.08, $[M+Na]^+$, (68)<br>266.06, $[M+2Na-H]^+$, (2)<br>465.17, $[2M+Na]^+$, (100)<br>487.15, $[2M+2Na-H]^+$, (1) | Mannose ($C_6H_{12}O_6$) | 180.0634 | 203.05, $[M+Na]^+$, (98)<br>225.03, $[M+2Na-H]^+$, (1.4)<br>383.11, $[2M+Na]^+$, (100)<br>405.10, $[2M+2Na-H]^+$, (4)<br>427.08, $[2M+3Na-2H]^+$, (0.2) |
| *N*-Acetyl-Neuraminic acid ($C_{11}H_{19}NO_9$) | 309.1060 | 332.09, $[M+Na]^+$, (16)<br>354.08, $[M+2Na-H]^+$, (100)<br>376.06, $[M+3Na-2H]^+$, (0.5)<br>641.20, $[2M+Na]^+$, (0.4)<br>663.18, $[2M+2Na-H]^+$, (2)<br>685.16, $[2M+3Na-2H]^+$, (28) | Fucose ($C_6H_{12}O_5$) | 164.0685 | 187.06, $[M+Na]^+$, (69)<br>209.04, $[M+2Na-H]^+$, (1)<br>351.13, $[2M+Na]^+$, (100) |
| Glucuronic acid ($C_6H_{10}O_7$) | 194.0427 | 195.04, $[M+H]^+$, (2)<br>217.03, $[M+Na]^+$, (6)<br>239.01, $[M+2Na-H]^+$, (100)<br>433.06, $[2M+2Na-H]^+$, (1)<br>455.04, $[2M+3Na-2H]^+$, (50) | D0A4 ($C_{14}H_{20}NNaO_{14}S$) | 481.0502 | 482.05, $[M+H]^+$, (1)<br>504.03, $[M+Na]^+$, (51)<br>526.02, $[M+2Na-H]^+$, (100) |
| ICA ($C_6H_9N_3OS$) | 171.0466 | 172.05, $[M+H]^+$, (10)<br>194.04, $[M+Na]^+$, (100)<br>216.02, $[M+2Na-H]^+$, (2)<br>365.08, $[2M+Na]^+$, (33) | D0A6 ($C_{14}H_{20}NNaO_{14}S$) | 481.0502 | 482.06, $[M+H]^+$, (2)<br>504.04, $[M+Na]^+$, (45)<br>526.02, $[M+2Na-H]^+$, (100) |

1. The relative Intensity (%) value of observed ions are given in parentheses next to each complex ion. The most intense peak in a single stage MS spectrum is defined as 100.

**Table S3.** Characteristic MS Peaks of 1:1 ICA-Carbohydrate complexes and their MS/MS products

(M denotes the corresponding carbohydrate molecule)

| Analytes | Observed m/z | MS/MS Product Ion |
|---|---|---|
| ICA + carbohydrate | Mass of adduct ion (Intensity, S/N) | Mass of Product ion, (Intensity) |
| Galactose | 374.10, [ICA+M+Na]$^+$, (41, 15417) | 194.04, [ICA+Na]$^+$, (100)<br>203.05, [M+Na]$^+$, (87) |
| Glucose | 374.10, [ICA+M+Na]$^+$, (0.9, 804) | 194.04, [ICA+Na]$^+$, (100)<br>203.05, [M+Na]$^+$, (8) |
| | 396.08, [ICA+M+2Na-H]$^+$, (0.2, 195) | 216.02, [ICA+2Na-H]$^+$, (51)<br>225.03, [M+2Na-H]$^+$, (100) |
| Galactosamine | 351.13, [ICA+M+H]$^+$, (55, 33497) | 162.08, [M-$H_2O$+H]$^+$, (13)<br>172.05, [ICA+H]$^+$, (19)<br>180.09, [M+H]$^+$, (100) |
| | 373.12, [ICA+M+Na]$^+$, (16, 10165) | 194.04, [ICA+Na]$^+$, (46)<br>202.07, [M+Na]$^+$, (100) |
| Glucosamine | 351.13, [ICA+M+H]$^+$, (43, 18890) | 162.08, [M-$H_2O$+H]$^+$, (10)<br>172.05, [ICA+H]$^+$, (53)<br>180.09, [M+H]$^+$, (100) |
| *N*-Acetyl-Galactosamine | 393.14, [ICA+M+H]$^+$, (0.3, 63) | 172.05, [ICA+H]$^+$, (68)<br>222.10, [M+H]$^+$, (100) |
| | 415.12, [ICA+M+Na]$^+$, (15, 3030) | 194.03, [ICA+Na]$^+$, (0.2)<br>244.08, [M+Na]$^+$, (100) |
| | 437.10, [ICA+M+2Na-H]$^+$, (0.5, 107) | 244.08, [M+Na]$^+$, (47)<br>266.06, [M+2Na-H]$^+$, (100) |
| *N*-Acetyl-Glucosamine | 415.13, [ICA+M+Na]$^+$, (53, 13720) | 194.04, [ICA+Na]$^+$, (12)<br>244.08, [M+Na]$^+$, (100) |
| *N*-Acetyl-Neuraminic acid | 503.12, [ICA+M+Na]$^+$, (0.03, 15) | 194.03, [ICA+Na]$^+$, (42)<br>332.08, [M+Na]$^+$, (39) |
| | 525.12, [ICA+M+2Na-H]$^+$, (3.1, 1569) | 354.08, [M+2Na-H]$^+$, (100) |
| | 547.10, [ICA+M+3Na-2H]$^+$, (0.03, 17) | 332.09, [M+Na]$^+$, (8)<br>354.08, [M+2Na-H]$^+$, (100) |
| Glucuronic acid | 366.08, [ICA+M+H]$^+$, (7, 2658) | 194.04, [ICA+Na]$^+$, (100)<br>195.04, [M+H]$^+$, (73) |
| | 388.08, [ICA+M+Na]$^+$, (0.1, 43) | 194.04, [ICA+Na]$^+$, (100) |

| Analytes | Observed m/z | MS/MS Product Ion |
|---|---|---|
| ICA + carbohydrate | Mass of adduct ion (Intensity, S/N) | Mass of Product ion, (Intensity) |
| α-$^M$Glu | 388.12, [ICA+M+Na]$^+$, (2.1, 1494) | 194.04, [ICA+Na]$^+$, (100)<br>217.07, [M+Na]$^+$, (15) |
| β-$^M$Glu | 366.08, [ICA+M+H]$^+$, (6, 1935) | 172.05, [ICA+H]$^+$, (3)<br>194.05, [ICA+Na]$^+$, (100)<br>195.04, [M+H]$^+$, (72) |
| | 388.12, [ICA+M+Na]$^+$, (65, 21034) | 194.04, [ICA+Na]$^+$, (100)<br>217.07, [M+Na]$^+$, (14) |
| Maltose | 514.17, [ICA+M+H]$^+$, (1.5, 451) | 172.06, [ICA+H]$^+$, (100) |
| | 536.15, [ICA+M+Na]$^+$, (9.5, 3093) | 194.04, [ICA+Na]$^+$, (2)<br>365.11, [M+Na]$^+$, (100) |
| Cellobiose | 514.17, [ICA+M+H]$^+$, (1.2, 838) | 172.06, [ICA+H]$^+$, (100) |
| | 536.15, [ICA+M+Na]$^+$, (1.1, 846) | 365.11, [M+Na]$^+$, (100) |
| Xylose | 344.09, [ICA+M+Na]$^+$, (9.0, 5451) | 173.04, [M+Na]$^+$, (10)<br>194.03, [ICA+Na]$^+$, (100)<br>195.04, [M+2Na-H]$^+$, (5) |
| | 366.08, [ICA+M+2Na-H]$^+$, (2.6, 1553) | 194.03, [ICA+Na]$^+$, (100)<br>195.04, [M+2Na-H]$^+$, (12) |
| | 388.06, [ICA+M+3Na-2H]$^+$, (0.1, 65) | 194.03, [ICA+Na]$^+$, (13)<br>216.02, [ICA+2Na-H]$^+$, (100) |
| Mannose | 374.10, [ICA+M+Na]$^+$, (20, 10478) | 194.03, [ICA+Na]$^+$, (78)<br>203.05, [M+Na]$^+$, (87) |
| | 396.08, [ICA+M+2Na-H]$^+$, (0.5, 291) | 216.02, [ICA+2Na-H]$^+$, (22)<br>225.03, [M+2Na-H]$^+$, (100) |
| Fucose | 358.10, [ICA+M+Na]$^+$, (20, 9978) | 187.06, [M+Na]$^+$, (4)<br>194.03, [ICA+Na]$^+$, (100) |
| | 380.08, [ICA+M+2Na-H]$^+$, (0.6, 290) | 194.03, [ICA+Na]$^+$, (9)<br>209.04, [M+2Na-H]$^+$, (31)<br>216.02, [ICA+2Na-H]$^+$, (100) |
| D0A4 | 675.09, [ICA+M+Na]$^+$, (0.01, 3) | 194.01, [ICA+Na]$^+$, (53)<br>504.03, [M+Na]$^+$, (100) |
| | 697.07, [ICA+M+2Na-H]$^+$, (0.1, 42) | 194.01, [ICA+Na]$^+$, (14)<br>526.02, [M+2Na-H]$^+$, (100) |
| D0A6 | 675.09, [ICA+M+Na]$^+$, (0.01, 11) | 194.04, [ICA+Na]$^+$, (14)<br>504.04, [M+Na]$^+$, (100) |
| | 697.07, [ICA+M+2Na-H]$^+$, (0.02, 22) | 194.04, [ICA+Na]$^+$, (16)<br>526.02, [M+2Na-H]$^+$, (100) |

The Relative Intensity (I%) and Signal to Noise Ratio (S/N) values are given in parentheses respectively next to each complex ion in observed m/z column. I% values are reported in parentheses next to each complex ion in MS/MS product ion column. The most intense peak is considered as 100.

**Table S4.** Characteristic MS Peaks of 2:1 ICA-Carbohydrate complexes and their MS/MS products

(M denotes the corresponding carbohydrate molecule)

| **Analyte** | **Observed m/z** | **MS/MS Product Ion** | **Analyte** | **Observed m/z** | **MS/MS Product Ion** |
|---|---|---|---|---|---|
| ICA + carbohydrate | Mass of adduct ion, (Intensity, S/N) | Mass of Product ion, (Intensity) | ICA + carbohydrate | Mass of adduct ion, (Intensity, S/N) | Mass of Product ion, (Intensity) |
| Galactose | 545.13, [2ICA+M+Na]$^+$, (0.1, 35) | 365.06, [2ICA+Na]$^+$, (87) | α-$^{M}$Glu | 559.08, [2ICA+M+Na]$^+$, (0.02, 14) | Not measured |
| Glucose | Not measured | | β-$^{M}$Glu | 559.14, [2ICA+M+Na]$^+$, (0.04, 12) | 365.06, [2ICA+Na]$^+$, (87) |
| Galactosamine | 522.16, [2ICA+M+H]$^+$, (0.1, 60) | 180.09, [M+H]$^+$, (5) | Maltose | Not measured | |
| Glucosamine | Not measured | | Cellobiose | Not measured | |
| *N*-Acetyl-Galactosamine | 586.15, [2ICA+M+Na]$^+$, (1.3, 264) | 244.08, [M+Na]$^+$, (12) | Xylose | 537.11, [2ICA+M+2Na-H]$^+$, (0.01, 4) | 194.03, [ICA+Na]$^+$, (52) |
| | 608.13, [2ICA+M+2Na-H]$^+$, (0.1, 11) | 216.02, [ICA+2Na-H]$^+$, (100)<br>244.08, [M+Na]$^+$, (97) | | 559.09, [2ICA+M+3Na-2H]$^+$, (0.03, 19) | 194.03, [ICA+Na]$^+$, (9)<br>216.02, [ICA+2Na-H]$^+$, (15) |
| *N*-Acetyl-Glucosamine | Not measured | | Mannose | 567.11, [2ICA+M+2Na-H]$^+$, (0.03, 18) | 216.02, [ICA+2Na-H]$^+$, (100) |
| *N*-Acetyl-Neuraminic acid | 674.17, [2ICA+M+Na]$^+$, (0.05, 24) | 332.08, [M+Na]$^+$, (100)<br>354.08, [M+2Na-H]$^+$, (50) | Fucose | 551.11, [2ICA+M+2Na-H]$^+$, (0.02, 12) | 194.04, [ICA+Na]$^+$, (33)<br>209.04, [M+2Na-H]$^+$, (16)<br>216.02, [ICA+2Na-H]$^+$, (100) |
| | 696.15, [2ICA+M+2Na-H]$^+$, (0.1, 53) | 354.08, [M+2Na-H]$^+$, (100) | | 573.11, [2ICA+M+3Na-2H]$^+$, (0.01, 9) | 216.02, [ICA+2Na-H]$^+$, (100) |
| Glucuronic acid | 559.11, [2ICA+M+Na]$^+$, (0.02, 6) | 194.03, [ICA+Na]$^+$, (15) | D0A4 | 868.13, [2ICA+M+2Na-H]$^+$, (0.02, 23) | 194.01, [ICA+Na]$^+$, (48)<br>526.02, [M+2Na-H]$^+$, (100) |
| | | | D0A6 | 868.12, [2ICA+M+2Na-H]$^+$, (0.01, 13) | 194.04, [ICA+Na]$^+$, (43)<br>526.02, [M+2Na-H]$^+$, (100) |

The Relative Intensity (I%) and Signal to Noise Ratio (S/N) values are given in parentheses respectively next to each complex ion in observed m/z column. I% values are reported in parentheses next to each complex ion in MS/MS product ion column. The most intense peak is considered as 100.

## 6. Preparation of STM probes and substrates

### 6.1 Functionalization of STM probes

The functionalization of STM probes followed a procedure developed in our lab.[2] First, the STM probes were made from a 0.25 mm Pd wire (California Fine Wires) by AC electrochemical etching in a mixed solution of 36% hydrochloric acid and ethanol (1:1), and then coated with a high-density polyethylene (HDPE) film, having an open apex with a few tens of nanometers in diameter. The insulated probes were gently cleaned by ethanol (200 proof), dried with a nitrogen flow, immersed in an ethanolic solution of ICA (0.5 mM, degassed by argon) for 20 hours at room temperature, and then gently rinsed with ethanol and dried with nitrogen. All the STM probes were freshly prepared before each experiment.

### 6.2 Functionalization and characterization of palladium substrates

The substrate was prepared by depositing 100 nm thick palladium on a 750 μm silicon wafer coated with a 10 nm titanium adhesion layer using electron-beam evaporator (Lesker PVD 75). Palladium substrates were functionalized with ICA in the same way as we did for the STM probes and characterized with various physical and chemical tools.

#### 6.2.1 Thickness

We characterized the substrate with ellipsometry. The bare palladium substrate was annealed by hydrogen flame and its thickness was measured using Gaertner L 123b Ellipsometer (Gaerner Scientific Corporation) prior to functionalization. A refractive index of 1.50 was assumed for the organic thin film.[3] The monolayer measured 9.10±0.41 Å in thickness on average from five measurements in different locations of two samples. The ICA molecule was estimated by ChemDraw 3D to be ~8.3 Å long.

#### 6.2.2 Contact angle

Static water contact angles were measured for hydrogen annealed palladium substrate prior to SAM formation and for functionalized substrate after SAM formation using an Easydrop Drop Shape Analysis System (KRÜSS GmbH, Hamburg). Volume of each water droplet for static contact angle measurements was 1 μL. 5-6 measurements were taken on different locations of each functionalized and bare palladium substrates. The contact angle for the bare palladium substrate was 8.3 ± 2.0°, for the ICA monolayer 33.1 ± 5.1°

#### 6.2.3 FTIR spectra

The FTIR spectra were recorded using a Nicolet 6700 FT-IR (Thermo Electron Corporation) equipped with a MCT detector at a 4 $cm^{-1}$ resolution in a range from 4000 to 650 $cm^{-1}$, with an attenuated total reflection accessory (Smart Orbit, Thermo Electron Corporation) for an ICA powder with 128 scans (3500-1000 $cm^{-1}$ shown in Figure S2, a) and with a surface grazing angle accessory (Smart SAGA, Thermo Electron Corporation) with 256 scans for the ICA monolayer [1] (3500-1000 $cm^{-1}$ shown in Figure S2, b). The ICA powder sample gave broad bands in the region of 3400-2800 $cm^{-1}$, indicating the intermolecular hydrogen-bonding interactions between the ICA molecules; in contrast, the ICA monolayer shows very sharp peaks in the same region due to the removal of the intermolecular hydrogen bonds. Both spectra show the vibrations of the amide function in the region of 1700-1600 $cm^{-1}$.

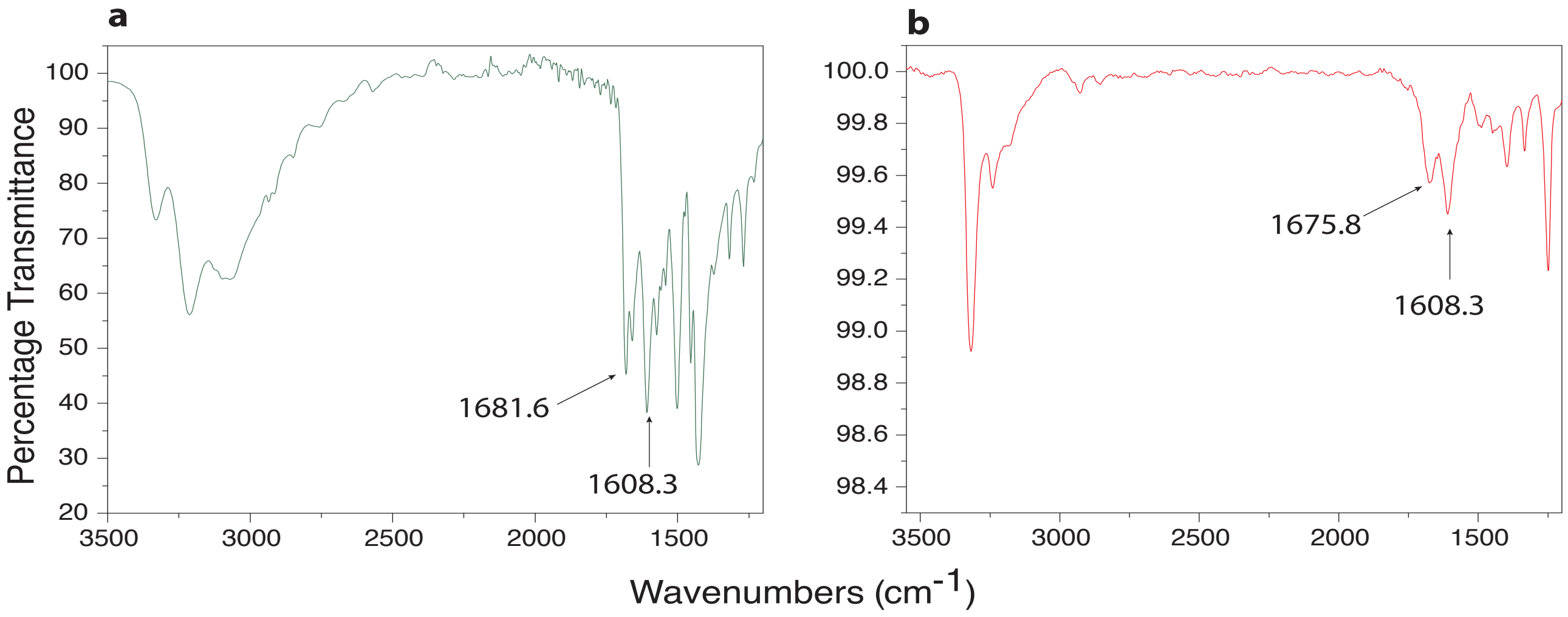

**Figure S2. FTIR spectra of (a) ICA powder and (b) ICA monolayer**

### 6.2.4 X-ray photoelectron spectroscopy (XPS)

XPS spectra were obtained using a VG ESCALAB 220i-XL photoelectron spectrometer and Al-Kα radiation (15keV) at 6 x 10E-10 mbar base pressure. The high resolution spectra of C(1s), Pd(3d), N(1s) and S(2p) were recorded at a pass energy of 20 eV and wide scan spectra were obtained at pass energy of 150 eV. The data analysis was carried out using CasaXPS software. C(1s), N(1s), and S(2p) core peaks were fitted and the ICA element ratio was calculated through area integral of peaks. Table S5 shows the found elemental ratio, which is close to the calculated ratio.

**Table S5.** Element compositions of the imidazole monolayer from XPS

| Element | Atomic Percentage (%) | Found elemental ratio | Calculated elemental ratio |
|---|---|---|---|
| S 2p | 5.17 | 1 | 1 |
| C 1s | 26.77 | 5.2 | 6 |
| N 1s | 12.22 | 2.4 | 3 |

### 7. STM experimental details and conditions

The RT measurements were carried out in PicoSPM (Agilent Technologies) with customized LabView interface for data acquisition. Each tip was tested to ensure that the current leakage was less than 1 pA in a PB solution at a 500 mV bias. The current set point was 4 pA, which corresponds to a gap size of about 2.5 nm between the tip and substrate.[4] The tip approached to substrate under 1.0 integral and proportional gain servo control. The surface first was scanned to ensure that the tip is not over-coated by high-density polyethylene (HDPE), so electrodes were a good condition for the RT measurement. After a clear grain structure on the Pd substrate was obtained, the tip was withdrawn about 1 μm and the bias was turned off to avoid possible damage to the ICA layer during the 2-hour instrument stabilization. Sequentially, the Tip was re-engaged, and the integral and proportional gains were set to 0.1. The response time was determined by noise spectrum (Power Spectral Density) under various gain values. Figure S4 shows that servo control distorts the signal. With 0.1 for the integral and proportional gain, spectrum under 30 Hz (corresponding 33 ms) was suppressed. The response time was long enough not to distort all spikes but some of long spikes. Tunneling current was Fourier transformed and plotted as a spectral density calculated by

$$PSD = \frac{2}{N \cdot \Delta t} \frac{Re^2 + Im^2}{f}$$

where $N = 50{,}000$ and $\Delta t = 20\ \mu s$.

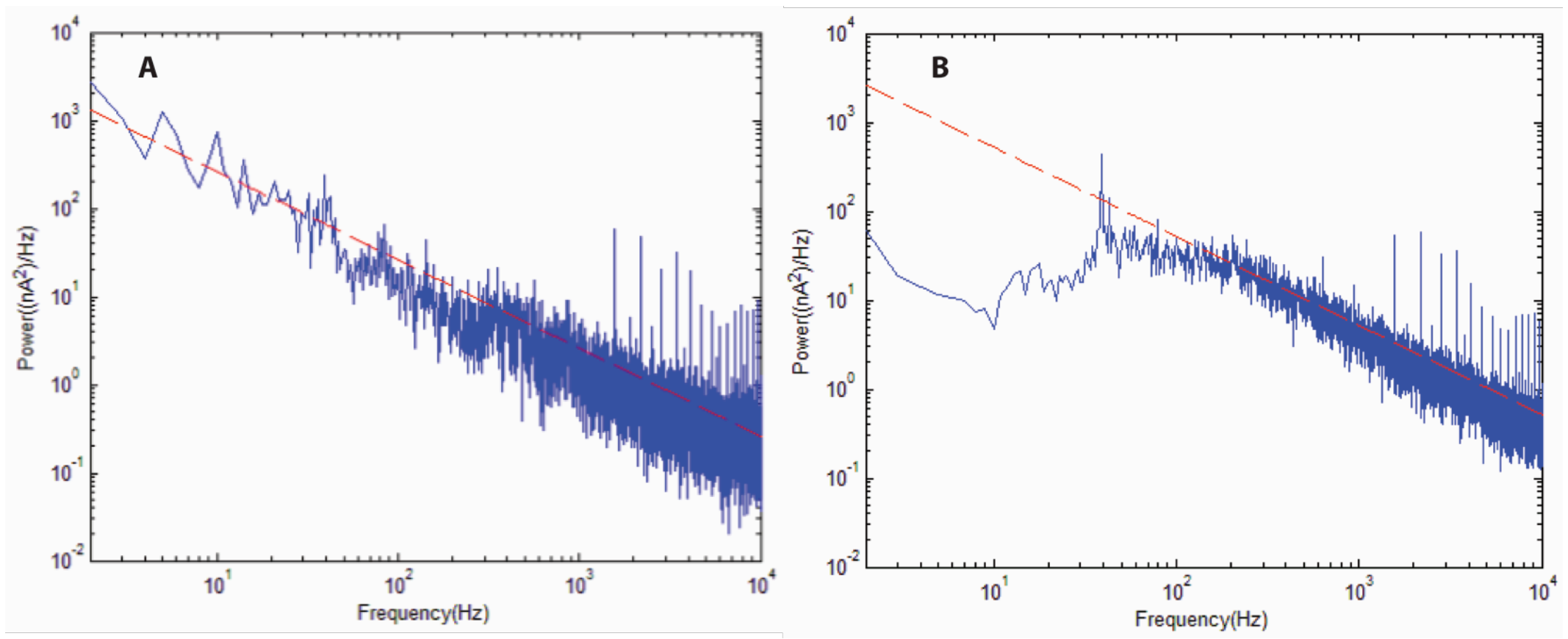

**Figure S3.** Noise spectrum (a) without servo control and (b) with servo control. Blue lines are the noise spectrum, and red lines are fits to 1/f spectrum.

### 7.1 Control experiments

As a routine, phosphate buffer solution (1 mM, pH 7.4) was used as a control for the RT measurement. Before every measurement, a spectrum of the buffer solution was recorded, only rarely manifesting noise spikes. Figure S4 illustrates the RT spectra measured with different electrodes in different solutions. Table S6 lists statistical data of electrical spike frequencies in different solutions with different tunnel gap.

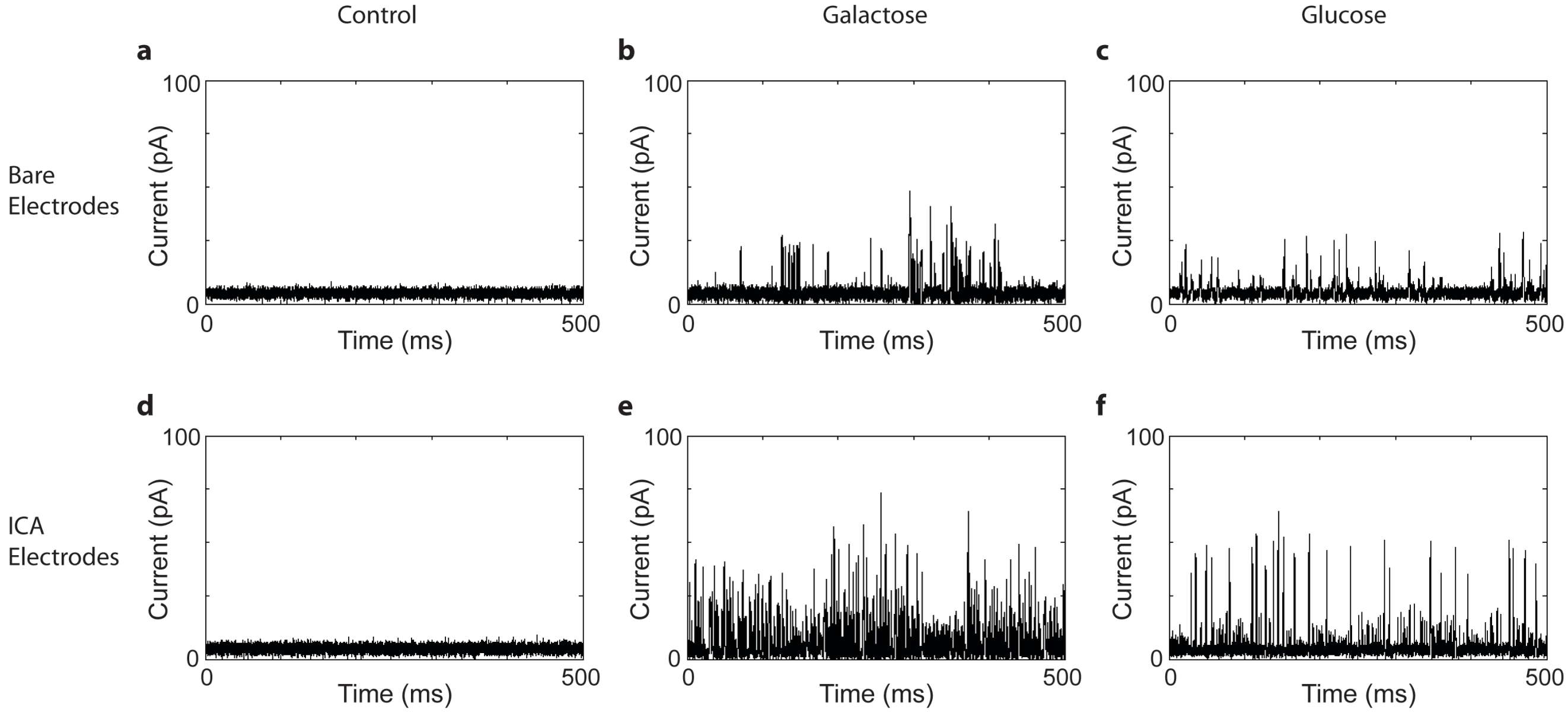

**Figure S4.** RT Spectra measured with bare electrodes (a, b, c) and with ICA functionalized electrodes (d, e, f) in different solutions

**Table S6.** Comparison of current spike frequency with different electrodes in phosphate buffer, Galactose and Glucose solution

| | Control | Galactose | Glucose |
|---|---|---|---|
| Bare electrodes | 0 | 0.33 | 0.34 |
| ICA-electrodes | 0 | 3.03 | 1.22 |

Unit: peaks/second

### 7.2 RT spectra of carbohydrates

Figure S5 shows exemplary RT spectra of all saccharides we have measured.

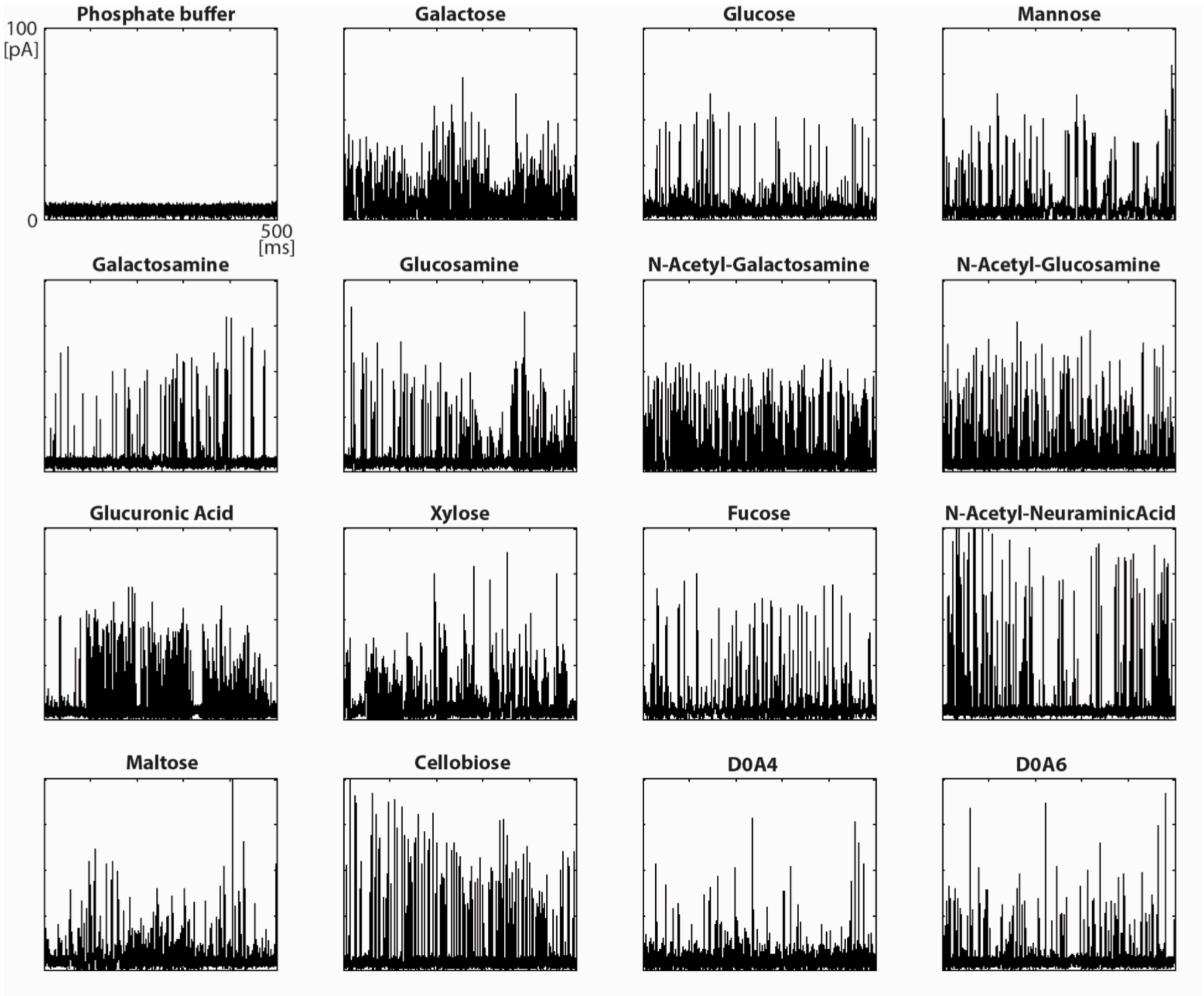

**Figure S5.** Representative RT Spectra of saccharides measured with ICA functionalized electrodes

## 8. RT signal frequency from α-MGlu as a function of concentration

The α-$^{M}$Glu was dissolved in the phosphate buffer to make a 1 mM stock solution, which was diluted to various concentrations from 500 μM to 100 pM. For each measurement, an analyte solution (200 μL) was injected into the liquid cell using a syringe attached to a micro filter. After the measurement, the liquid cell and electrodes were rinsed with the phosphate buffer solution (3 mL) through the fluidic channels to obtain a clean control signal. A pair of electrodes was able to carry out three measurements with different concentrations from lower to higher concentration and the measurement at each concentration was repeated at least 3 times (Table S7). The isotherm absorption data were analyzed in software OriginPro 2016 using the Levenberg-Marquardt algorithm for fitting to a Langmuir equation: $f(x) = a*(b \cdot x)/(1+b \cdot x)$

**Table S7.** RT signals frequency of α-$^{M}$Glu at different concentrations

| | Average peak counts (peaks/sec) | | | | | | | | |
|---|---|---|---|---|---|---|---|---|---|
| Conc. | Runs | | | | | Average | STD-mean | Normalized | STD-mean |
| 100 pM | 0.02 | 0.08 | 0.05 | | | 0.050 | 0.017 | 0.000 | 0.008 |
| 1 nM | 0.04 | 0.13 | 0.17 | | | 0.113 | 0.038 | 0.030 | 0.018 |
| 10 nM | 0.12 | 0.1 | 0.13 | 0.16 | 0.22 | 0.146 | 0.021 | 0.046 | 0.010 |
| 100 nM | 0.33 | 0.35 | 0.57 | 0.41 | | 0.415 | 0.054 | 0.174 | 0.026 |
| 1 μM | 0.86 | 1.17 | 0.88 | | | 0.970 | 0.100 | 0.437 | 0.048 |
| 10 μM | 1.89 | 1.61 | 1.78 | | | 1.760 | 0.081 | 0.813 | 0.039 |
| 100 μM | 2.33 | 1.99 | 1.99 | 1.75 | | 2.015 | 0.119 | 0.934 | 0.057 |
| 250 μM | 2.35 | 2.08 | 2.2 | 1.7 | | 2.083 | 0.139 | 0.966 | 0.066 |
| 500 μM | 2.26 | 1.8 | 2.4 | | | 2.153 | 0.181 | 1.000 | 0.086 |

## 9. Surface Plasmon Resonace (SPR) experiments

A gold chip was immersed into an absolute ethanol solution of ICA (100 μM) for 24 h, followed by rinsing with absolute ethanol and drying with a nitrogen flow, and used immediately. The instrument Bi 2000 from Biosensing Instrument was used for SPR measurements. An ICA modified gold chip was mounted on the instrument and calibrated with 1% ethanol in a PBS buffer, pH 7.4. A solution of α -$^{M}$Glu (500 μM) was flowed onto the chip at a rate of 50 μl/min over a period of 1.5 min, followed by flowing the PBS buffer (Figure S2). Association ($k_{on}$) and dissociation rate constants ($k_{off}$) were determined using built-in Biosensing Instrument SPR data analysis software version 2.4.6 (Table S5).

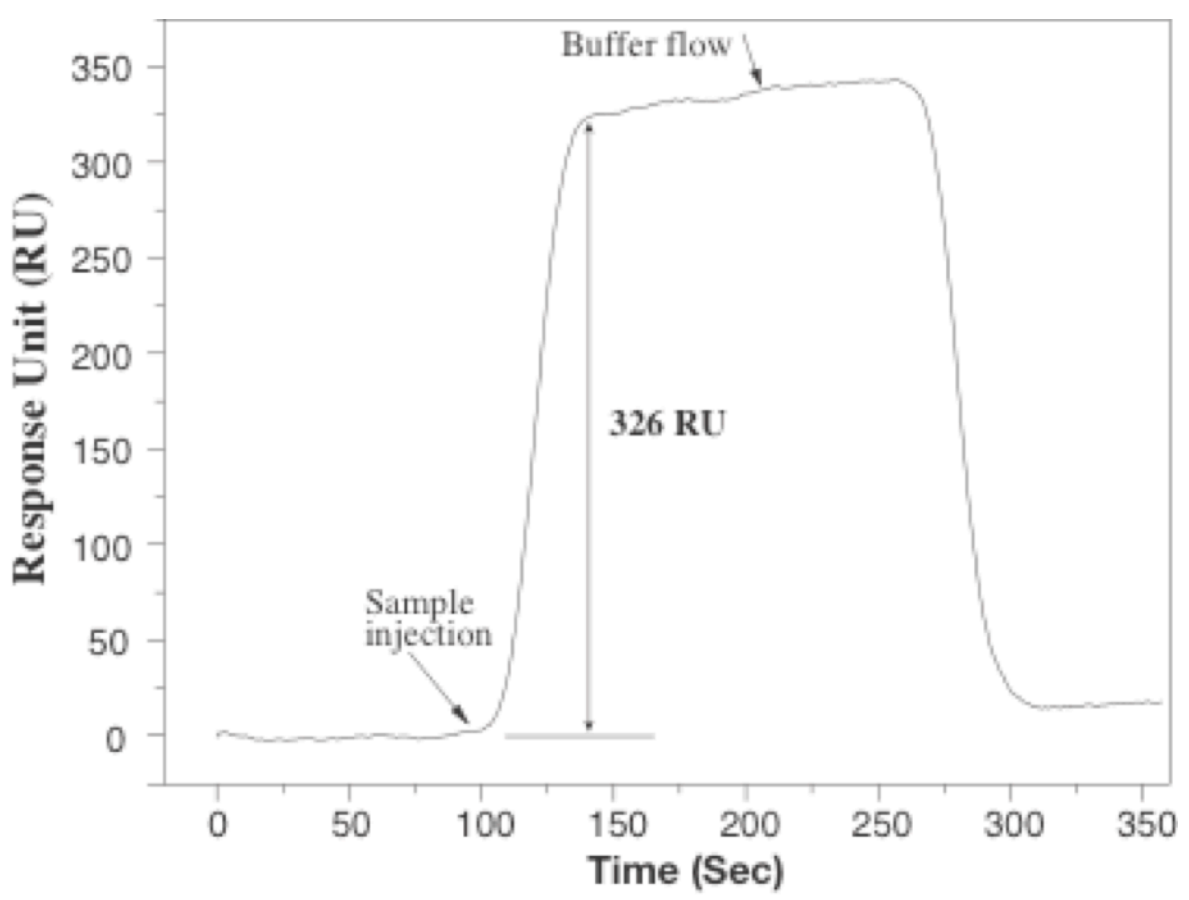

**Figure S6.** SPR sensorgram of adsorption and desorption of α-$^{M}$Glu on the ICA monolayer

**Table S8.** Kinetic parameters of α-$^{M}$Glu on the ICA monolayer[1]

| α -$^{M}$Glu | $k_{on}$ ($M^{-1}$ $s^{-1}$) | $k_{off}$ ($s^{-1}$) | $K_d$[2] (mM) | Res.sd |
|---|---|---|---|---|
| 500 uM | 23.05±1.34 | 0.09±0.01 | 4.00±0.14 | 10.2 |

1. Each datum listed is an average of two measurements.
2. $K_d = k_{off}/k_{on}$

## References

1 Pramanik, B. N., Bartner, P. L., Mirza, U. A., Liu, Y.-H. & Ganguly, A. K. Electrospray ionization mass spectrometry for the study of non-covalent complexes: an emerging technology. *J. Mass Spectrom.* **33**, 911-920 (1998).

2 Tuchband, M., He, J., Huang, S. & Lindsay, S. Insulated gold scanning tunneling microscopy probes for recognition tunneling in an aqueous environment. *Rev. Sci. Instrum.* **83**, 015102 (2012).

3 Ulman, A. *An Introduction to Ultrathin Organic Films: From Langmuir--Blodgett to Self--Assembly*. (Academic Press, 1991).

4 Chang, S., He, J., Zhang, P., Gyarfas, B. & Lindsay, S. Gap distance and interactions in a molecular tunnel junction. *J. Am. Chem. Soc.* **133**, 14267-14269 (2011).